\documentclass[%
 reprint,
nofootinbib,
 amsmath,amssymb,
 aps,
]{revtex4-1}

\usepackage{graphicx}
\usepackage{amssymb}
\usepackage{dcolumn}
\usepackage{bm}

\usepackage{lineno}

\def\ba{\begin{array}}
\def\ea{\end{array}} 
\def\bea{\begin{eqnarray}}
\def\eea{\end{eqnarray}}
\def\beq{\begin{equation}}
\def\eeq{\end{equation}}
\def\ben{\begin{enumerate}}
\def\een{\end{enumerate}}

\def\brr{\begin{array}}
\def\err{\end{array}}

\def\calC{\S}

\begin{document}


\title{The cosmological constant 
and the size of the Causal Universe}



\author{Enrique Gazta\~naga}
 \affiliation{
Institute of Space Sciences (ICE, CSIC), 08193 Barcelona, Spain \\
Institut d’Estudis Espacials de Catalunya (IEEC), 08034 Barcelona, Spain
}

\date{\today}

\begin{abstract}

The cosmological constant $\Lambda$ is a free parameter in Einstein's equations of gravity. We propose to fix its value with a boundary condition: test particles should be free when outside causal contact, e.g. at infinity. 
Under this condition, we show that constant vacuum energy does not change cosmic expansion and 
there can not be cosmic acceleration for an infinitely large and uniform Universe.
The observed  acceleration  requires either 
a large Universe with evolving Dark Energy (DE) and equation of state $\omega>-1$ or a finite causal boundary (that we call Causal Universe) without DE.  The former can't explain why $\Omega_\Lambda \simeq 2.3 \Omega_m$ today, something that comes naturally with a finite Causal Universe. This boundary condition, combined with the anomalous lack of correlations  observed above 60 degrees in the CMB predicts  
$\Omega_\Lambda \simeq 0.70$ for a flat universe, with independence of any other measurements. This solution  provides new clues and evidence for inflation and removes the need for Dark Energy or Modified Gravity.

\end{abstract}


\maketitle


\section{Introduction}
\label{S:1}
 Measurements of cosmic expansion (see e.g. \cite{Planck2018,des2018,Tutusaus2017,Gaztanaga2009,Gaztanaga2006})  point to a model with $\Lambda$, that we refer to as $\Lambda$CDM.
For a flat homogeneous metric $ds^2=a^2( d\eta^2 - d\chi^2)$, the standard
expanding equations are:
\bea
H(a)^2 &\equiv& \left({\dot{a}\over{a}}\right)^2 
\label{eq:Hubble}
=  {8\pi G\over{3}} \rho(a) +  \Lambda
\\ \nonumber
\rho(a) &=& \rho_m a^{-3}+\rho_r a^{-4} +\rho_{\rm vac}
\eea
and its derivative.  Where $\rho_m$ is the pressureless matter density today ($a=1$),
$\rho_r$ corresponds to radiation (with pressure $p_r=\rho_r /3$)  and $\rho_{\rm vac}$  represents vacuum energy ($p_{vac}= -\rho_{\rm vac} $). How this equation emerge (or can be valid) in a Universe of finite age and therefore a finite causal size? How can $\rho(a)$ be the same everywhere at a fix $a$? We will argue here that this problem relates to cosmic acceleration.
Both  $\rho_{\rm vac}$ and $ \Lambda$ produce cosmic acceleration but we can only measure the combination:
\beq
\rho_{\Lambda}  \equiv  \rho_{\rm vac} + \frac{\Lambda}{8\pi G}  .
\label{eq:rhoHlambda}
\eeq
The measured $\rho_\Lambda$ is extremely small compared to what we expect for $\rho_{\rm vac}$. Moreover, $\rho_\Lambda \simeq 2.3\rho_m$, which is a remarkable and puzzling coincidence.
Possible solutions are:  I) $\Lambda=0$, II)  $\rho_{\rm vac}=0$ or III) a cancellation between them (for a review see \cite{Weinberg1989,Elizalde1990, Carroll,Huterer,Martin12,Lobo2001,Gaztanaga2002,Lue}). 
In option I)  the observed $\rho_\Lambda$ originates only from $\rho_{\rm vac}$ or dark energy (DE).
But in quantum field theory (QFT) $\rho_{vac}=\infty$  and observables  depend only on energy differences. If this is also true for gravity, only $\Lambda$ contributes to $\rho_\Lambda$. This is option II), which includes modified gravity models. 

In this paper we take $\Lambda$ to be a fundamental part of the gravitational interactions. 
To understand the role of  $\Lambda$  consider first the implications in classical physics.
This corresponds to adding a Hooke term, i.e. proportional to distance, to the gravitational acceleration:
\beq
\vec{g} = - \left( \frac{G m}{r^3} - \frac{\Lambda}{3} \right) \vec{r}
\label{eq:hooke}
\eeq
or equivalently a Poisson equation $\nabla^2 \phi = 4\pi G \rho -\Lambda$ (see Eq.\ref{poisson2}).
This generalization retains  a key property required for gravity, that a spherical mass shell of arbitrary density produces a gravitational field which is identical to a point source of equal mass in its center \cite{Wilkins,CalderLahav2008}. 
Here, in addition, we require that test  particles should be free when outside causal contact. 
If we request $\vec{g} \Rightarrow 0$ for $r \Rightarrow \infty$ in the equation above, we obtain $\Lambda=0$. The observational fact that $\Lambda >0$ is therefore indicating that  $r  \Rightarrow r_\calC < \infty$ (in agreement with the finite age of the Universe) and gives $\Lambda= 4\pi G \rho(r<r_\calC)$, which is related to the coincidence and horizon problems.

Particles separated by distances larger than the comoving Hubble radius 
$d_H(t)=c/[a(t)H(t)]$ can't communicate at time $t$. Distances larger than the horizon
\beq
\eta(a) = c \int_{0}^{t}  \frac{dt}{a(t)} = c \int_{0}^{a}  {d\ln(a)} ~d_H(a) ,
\label{eq:eta}
\eeq
have never communicated.
This either means that the initial conditions where acausally  smooth to start with or that there is a mechanism like inflation \cite{Dodelson,Liddle1999,Brandenberger,Martin} which inflates causally connected regions outside the Hubble radius.
 This  allows the full observable Universe to originate from a very small causally connected  homogeneous patch, which here we call the Causal Universe,  $\chi_\calC$. 
During inflation, $d_H$  decreases  which freezes out communication on comoving scales larger than the horizon  $\chi_\calC \simeq \eta(a_i)=d_H(t_i)$ when inflation begins, at $a_i=a(t_i)$. 
When inflation ends, radiation from reheating makes $d_H$ grow again. When  $\chi_\calC$ re-enters causal contact, we will see that the Universe starts another inflationary epoch so that  $\chi_\calC$  keeps frozen.  Thus, causality  can only play a role for comoving scales $\chi<\chi_\calC$.  The Causal Universe $\chi_\calC$ is therefore fixed in comoving coordinates and is the same for all times, while the horizon $\eta$ and $d_H$ change with time. Fig.\ref{Fig:horizon} illustrates this situation. We conclude that Eq.\ref{eq:Hubble}  only makes sense for comoving scales $\chi<\chi_\calC$.

\section{Fixing the value of $\Lambda$}
\label{sec:infinite}

The symmetries of Einstein's gravitational field equations allow a cosmological constant $\Lambda$ (\cite{Weinberg1972}):

\beq
R_{\mu\nu} + \Lambda g_{\mu\nu} = -8\pi~G~(T_{\mu\nu}-\frac{1}{2} g_{\mu\nu} T^\sigma_\sigma) ,
\label{eq:rmunu}
\eeq

For 
a homogeneous and isotropic perfect relativistic fluid with  density $\bar{\rho}$ and pressure  $\bar{p} \equiv \omega \bar{\rho}$:

\beq
T_{\mu\nu} = \bar{p} g_{\mu\nu} + (\bar{p}+\bar{\rho}) u_\mu u_\nu 
\eeq

As explained in the introduction, causality can only be efficient for $\chi<\chi_\calC$.
How can we implement this condition in $T_{\mu\nu}$?
Larger scales can have no effect on the metric, which is equivalent to say that $T_{\mu\nu}$
becomes zero for $\chi> \chi_\calC$. This corresponds to:
\beq
\bar{\rho}=\rho ~{\cal{H}}(\chi_\calC-\chi) 
\label{eq:Heaviside}
\eeq
where ${\cal{H}}$ is the Heaviside step function and $\rho$ is the mean density inside $\chi_\calC$. This does not mean that space is  empty for $\chi>\chi_\calC$, but just that it can have no effect on the metric as seem by our observer.
An observer situated close to the causal boundary of our first observer
will find a similar solution, but could measure different values for $\rho$, $p$ and $\Lambda$ depending on the initial conditions. The solutions from different acausal regions could be matched \cite{Sanghai-Clifton} which creates a smooth but inhomogeneous background across disconnected regions with  an infrared cutoff in the spectrum of homogeneities for $\chi>\chi_\calC$.

On scales $\chi<\chi_\calC$ we  have a homogeneous expanding universe. On larger scales we look for a solution as a perturbation around Minkowski space. In the weak field limit,  $g_{00} \simeq - (1+ 2\phi)$, where $\phi$ is the Newtonian  potential (see eg
\cite{Weinberg1972}). The field equations become a covariant generalization of Poisson equation:

\beq
R_{00} \simeq -  \partial^\mu \partial_\mu \phi = 
- 4\pi G  ~(\bar{\rho}+ 3 \bar{p})  + \Lambda ,
\label{poisson2}
\eeq
This is the same as Eq.6.13 in \cite{Peebles}, keeping
the time derivative term to have a covariant 4D d'Alambert operator. 
 We can use Stokes theorem 
to estimate the invariant Gauss flux of the 4D acceleration $\mathrm{g}_\mu \equiv \partial_\mu \phi$ 
in a 3D hyper-surface $\partial M$ of a 4D volume $M$:
\beq
\Phi= \oint_{\partial M} ~dx_\mu ~ \mathrm{g}^\mu  = 
- \int_M  d^4x \left[ 4\pi G  ~(\bar{\rho}+ 3 \bar{p})  - \Lambda  \right] ,
\label{eq:flux0}
\eeq
where $d^4x$ and $dx_\mu$ are the invariant 4D volume element and normal surface element of $\partial M$. 

\subsection{Causal Boundary condition}
\label{sec:boundary}

We require next that a test particle should be free (i.e. the metric is Minkowski)
outside causal contact.  Thus $\mathrm{g}_\mu=0$ and $\Phi=0$ for  $\chi>\chi_\calC$. This fixes $\Lambda$:

\beq
\Phi=0 \Rightarrow
\frac{\Lambda}{8\pi G} =   \frac{1}{2M_{\calC}} \int_{M_{\calC}}  d^4x  ~  (\rho+ 3p) ,
\label{eq:rhoH}
\eeq
where $M_{\calC}$ is the volume inside the lightcone to the surface $\partial M_{\calC}$ and we have use Eq.\ref{eq:Heaviside}.
As we approach the boundary  $\partial M_{\calC}$ there is no gravitational field 
and there is no energy associated with it.  Because of energy conservation, the $\Lambda$ term has to be constant and its value is the same at any cosmic time for our arbitrary observer.

\begin{figure}
\centering\includegraphics[width=1.\linewidth]{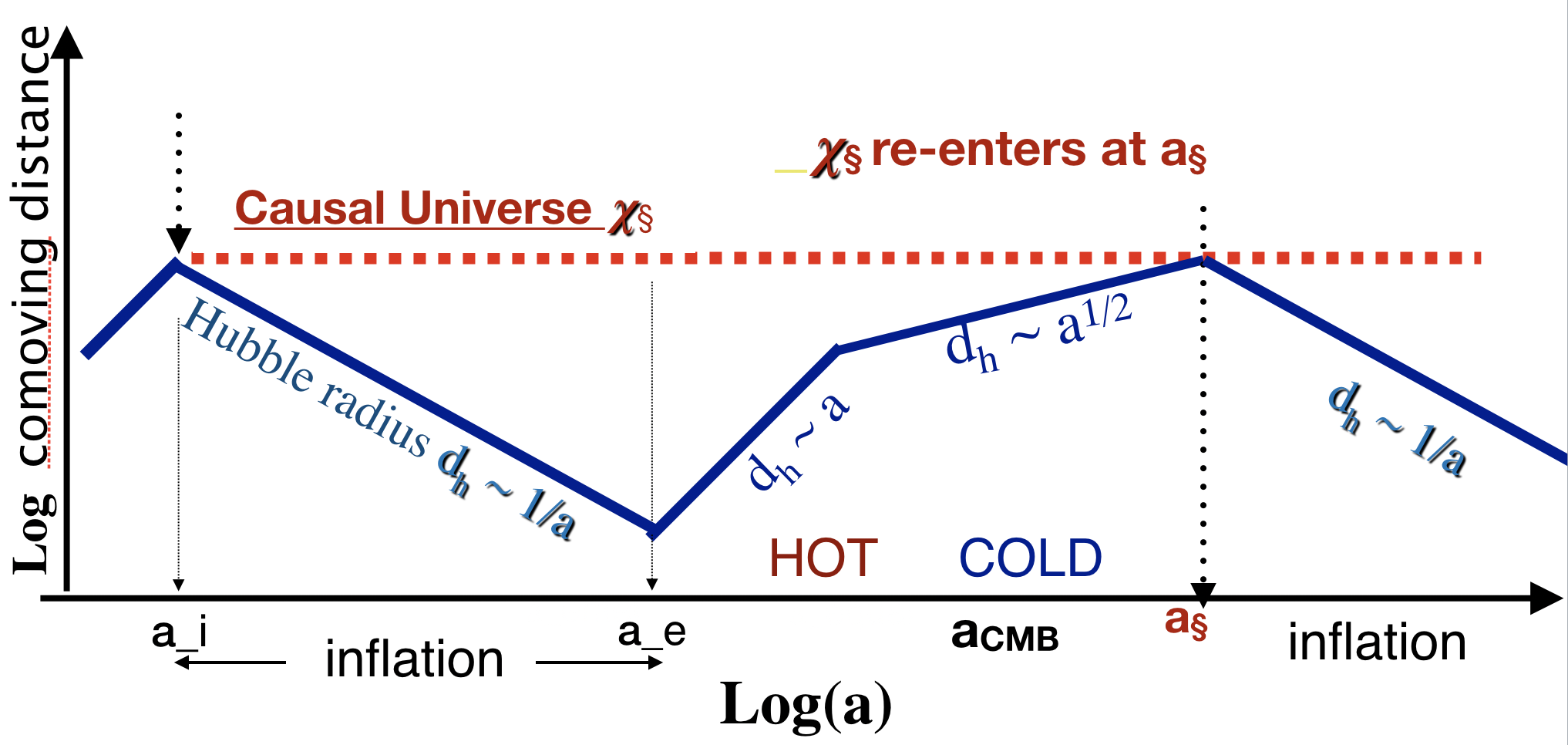}
\caption{Comoving 
Hubble radius  $d_H = c/(aH) $ (blue line) as a function of the scale factor $a$.  The Causal Universe $\chi_\S$ is identified with the region inside the largest causally connected scale at the beginning of inflation (red dashed line). 
During inflation  $d_H$ decreases so scales $\chi>d_H$ can no longer communicate. After inflation $d_H$ grows again (during HOT=radiation and COLD=matter domination) and  $\chi_\calC$ re-enters $d_H$ at some time $a_\calC$, which creates another inflation.}
\label{Fig:horizon}
\end{figure}

\subsection{Vacuum Energy does not gravitate}
\label{sec:vac}
Inside $\chi<\chi_\calC$, we can use Eq.\ref{eq:Hubble}
with $\rho= \rho_m+\rho_r+\rho_{\rm vac}$ and $p= \rho_r/3-\rho_{\rm vac}$, so that 
we can write Eq.\ref{eq:rhoH}  as:
\beq
\frac{\Lambda}{8\pi G} = \frac{\rho_m(\calC)}{2} + \rho_r(\calC) - \rho_{\rm vac}
\equiv \rho_{\calC} - \rho_{\rm vac} ,
\label{eq:m2}
\eeq
where $\rho_{\calC}$ is the matter and radiation contribution in the integral of Eq.\ref{eq:rhoH}. The values of $\rho_m$ and $\rho_r$ evolve with space-time, so that $\rho_{\calC}$ is the average contribution inside the volume $M_\calC$, while the vacuum density contribution is constant. 
As pointed out in Eq.\ref{eq:rhoHlambda}, $\rho_{\rm vac}$ has the same effect in  Eq.\ref{eq:rmunu} as $\Lambda$, and the only observable  is:

\beq
\rho_{\Lambda}  = \frac{\Lambda}{8\pi G}  + \rho_{\rm vac} =  {\rho}_{\calC} ,
\label{eq:rhoH2}
\eeq
where in the last equality we have used our causality condition in Eq.\ref{eq:m2}. So we see how vacuum energy cancels out and can not change the observed value of $\rho_{\Lambda}$, even for $\rho_{\rm vac}= \infty$, as predict by QFT.  
If  vacuum energy suffers a phase transition or changes in some other way, as  is believed to have happened during inflation, then this cancellation will not necessarily happen and $\rho_{\rm vac}$ could contribute to the effective value of $\rho_{\Lambda}$.

\subsection{Effective Dark Energy (DE)}
\label{sec:DE}

The general case considered here  is:
\bea
\rho_{DE}(a)  &=& \rho_{\rm vac} + \rho_{DE} ~a^{-3(1+\omega)}
\label{eq:DE}
\\ \nonumber
p_{DE}(a)  &=& -\rho_{\rm vac} + \omega ~\rho_{DE} ~ a^{-3(1+\omega)} ,
\eea
where only one component of DE is evolving.
We  then have from Eq.\ref{eq:rhoH} and Eq.\ref{eq:rhoHlambda}:
\beq
\rho_{\Lambda} 
= {\rho}_{\calC} +
 \rho_{DE} [1 +  \frac{1+3\omega}{2} \hat{a}_{\calC}^{-3(1+\omega)}] .
 \label{eq:rhoH3b}
\eeq
where $\hat{a}_{\calC}$ is some mean value of $a$ in the past light-cone of $a_\calC$ in Eq.\ref{eq:rhoH}.
This reduces to  $\rho_\Lambda = \rho_\calC$ for $\omega=-1$, which indicates that we need a finite $\chi_\calC$ to explain cosmic acceleration.
For $a_\calC \Rightarrow \infty$ 
we have $\rho_\calC \Rightarrow 0$  because  $\rho_m(a)$ and $\rho_r(a)$ tend to zero as we increase $a_\calC$. The same happens with $\hat{a}_{\calC}^{-3(1+\omega)}$ for $\omega >-1$, so that:
\beq
\rho_\Lambda  = \rho_{DE}   ~~ \rm{for}~~ {a}_\calC \Rightarrow \infty  ~~\rm{\&} ~~ \omega>-1 .
\label{eq:rhoDE}
\eeq
So evolving DE  could produce the observed cosmic acceleration in an infinitely large Universe. 
This solution does not explain why $\rho_{\Lambda} = \rho_{DE} \simeq 2.3\rho_m$. The original motivation to introduce DE was to understand why the vacuum energy $\rho_{\rm vac}$ can be as small as the measured $\rho_\Lambda$ \cite{Weinberg1989,Huterer,Martin12}. The causal boundary condition shows that $\rho_{\rm vac}$ does not contribute to $\rho_\Lambda$, which removes the motivation to have DE. So we will explore here a different way to get $\Omega_\Lambda \simeq 0.70$ without DE, i.e. for $\omega=-1$.

\begin{figure}[t]
\centering\includegraphics[width=1.0\linewidth]{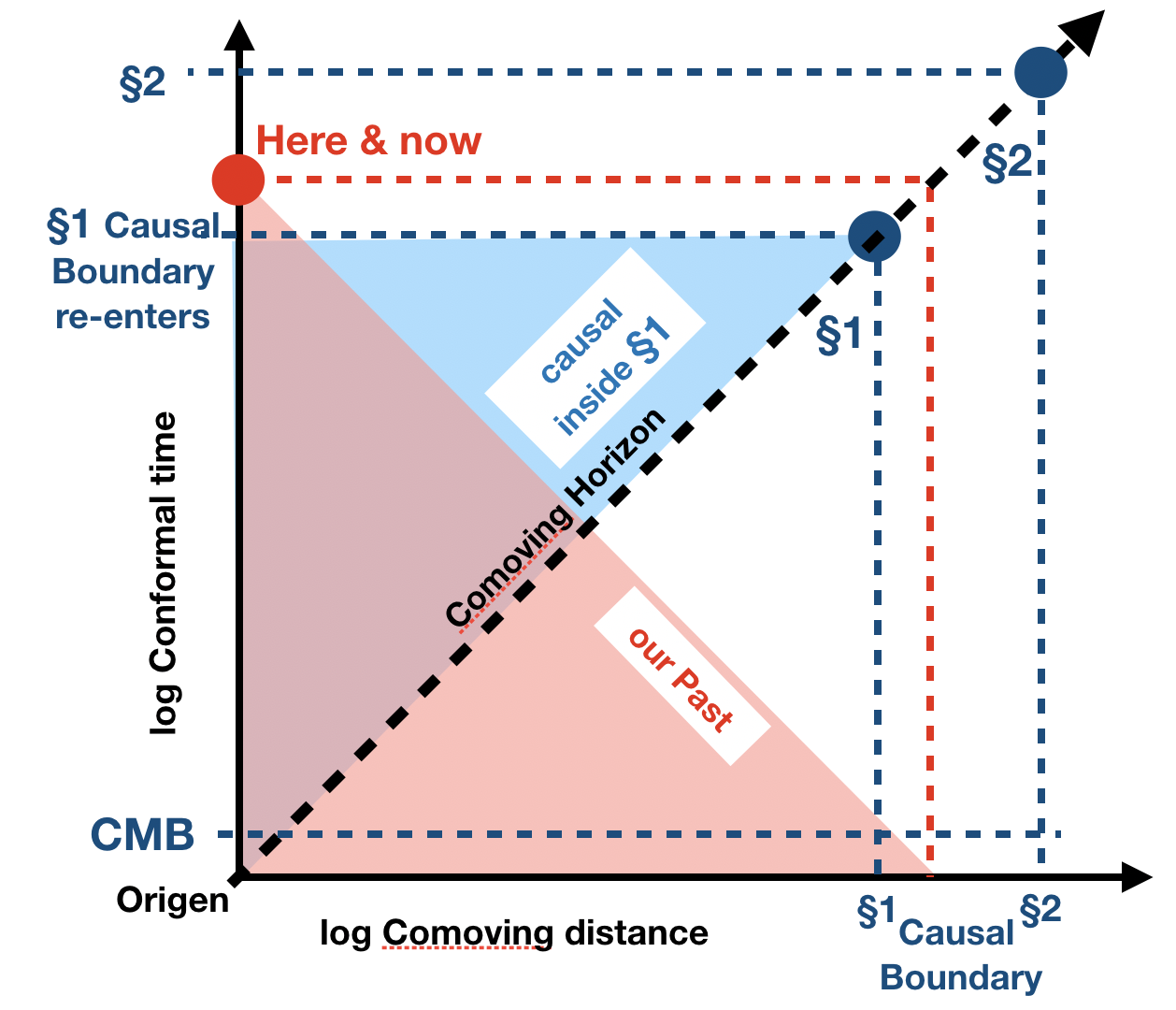}
\caption{Space-time diagram $ds^2=a^2( d\eta^2 - d\chi^2)$
for an expanding universe after inflation with two options (labeled $\S 1$ and  $\S 2$). The causal boundary (blue dot)  re-enters the Horizon some time that can be earlier ($\S 1$) or later  ($\S 2$) than our present time (red dot). Note how in case $\S 1$ we can observed regions outside the Causal Universe at the time when CMB was produce (see section \ref{sec:CMB}).}
\label{Fig:conformal}
\end{figure}

\section{The size of the Causal Universe}
\label{sec:size}

We  assume in this section that vacuum energy is constant after inflation ($\omega=-1$). In this case Eq.\ref{eq:rhoH} gives:
\beq
\rho_{\Lambda} = \rho_{\calC} =
 \frac{1}{2M_{\calC} } \int_{M_{\calC}}  d^4x  (\rho_m + 2\rho_r) .
\label{eq:rhoH4}
\eeq

From Eq.\ref{eq:eta}, the horizon after inflation 
is:
\beq
\chi(a) = \eta(a) - \eta(a_e)
\label{eq:chia}
\eeq
where $a_e$ represents the end of inflation. We then have $\chi_\calC = \chi(a_\calC) = \eta(a_i)$
where $a_{\calC}$ is the time when the causal boundary enters the horizon after inflation and $a_i$ the begining of inflation. Fig.\ref{Fig:horizon}-\ref{Fig:conformal} illustrate this.
We calculate $\rho_{\calC}$  in Eq.\ref{eq:rhoH4} by integrating  within the light-cone  of $\chi_{\calC}$:

\beq
\rho_{\calC}  = 
~ \frac{\int_{0}^{\chi_{\calC}}  d\chi ~ \chi^2 ~{a^3}~(\rho_m a^{-3} + 2\rho_r a^{-4})}{2 \int_{0}^{\chi_{\calC}}  d\chi  ~\chi^2 ~a^3} ,
\label{Eq:rhoHchi}
\eeq
where $a=a(\chi)$ in Eq.\ref{eq:chia}.
 For $H(a)$ we use Eq.\ref{eq:Hubble}
with $\Omega \equiv \rho/\rho_c$ 
$\rho_c=  3H_0^2 / 8\pi G$, 
$\Omega_r=4.2 \times 10^{-5}$ \cite{Planck2018}
and flat Universe
$\Omega_m=1-\Omega_\Lambda-\Omega_r$. 
We use Eq.\ref{Eq:rhoHchi} to solve $\Omega_\Lambda=\Omega_{\calC}$ numerically 
for $\Omega_\Lambda=0.69 \pm 0.01$ \cite{Planck2018}:

\bea
\chi_{\calC} &=& \left( 3.149 \pm 0.004 \right) \frac{c}{H_0} 
\label{eq:chi_H} \\
a_{\calC} &=& 0.933 \pm 0.004 .
\label{eq:a_H}
\eea
This scale factor corresponds to an age:

\begin{equation}
t_{\calC}= \left( 0.887 \pm 0.004 \right) \frac{1}{H_0} \simeq 12.5 {\rm Gyr} ,
\label{eq:age}
\end{equation}
compared to $t_{age} \simeq 0.955/H_0$ today,
i.e. about $\Delta t \simeq 0.84$ Gyr into our past. We can't observe this boundary $\chi_\calC$ today (see Fig.\ref{Fig:conformal}) but we will be able to observe it in the future and in our past (see section \ref{sec:CMB}).

\subsection{Inflation and the coincidence problem}
\label{sec:coincidence}

Eq.\ref{eq:rhoH4} indicates that when the causal boundary re-enters the Horizon the expansion becomes dominated by $\rho_\Lambda$. This is because $\rho_m(a_\calC)<\rho_\calC=\rho_\Lambda$, as density decreases with the expansion. This results in another inflationary epoch at $a=a_\calC$ which keeps the Causal Universe frozen (see Fig.\ref{Fig:horizon}).  
 We can now recast the coincidence problem (why $\rho_{\Lambda} \simeq 2.3 \rho_m$?)
into a new question: why we live at a time which is close to  $a_\calC$? 
 Looking  at Fig.\ref{Fig:horizon}, we can see that the best time to host observers
 is a time close to $a_\calC$ as the Universe is dominated by 
 $\rho_m$ (so there are galaxies) and the Hubble radius is the largest. 
There is nothing too special about this coincidence.

  The reason why $\chi_\calC \sim 3c/H_0$ and not some other value could reside in the details of inflation: when inflation begins $a_i$ and ends $a_e$ (see Fig.\ref{Fig:horizon}).
This recasts the coincidence problem into an opportunity to better understand inflation and the origin of homogeneity. We propose to identify $\chi_\calC=\eta(a_i)$ with the comoving  horizon before inflation begins at time $t_i$, $H_i=H(t_i)$ or $a_i=a(t_i)$:

\begin{equation}
a_i H_i =  c\chi_{\calC}^{-1} \simeq \left( 0.321 \pm 0.004\right) H_0
\label{eq:aH}
\end{equation}
The Hubble rate during inflation $H_I$ is proportional to the energy of inflation. During reheating this energy is converted into radiation:
$H_I^2 \simeq \Omega_r ~H_0^2 ~ a_e^{-4}$, with
$a_e \equiv e^{N} a_i$. We can combine  with Eq.\ref{eq:aH} to find:

\beq
a_i \chi_{\calC} = \frac{H_i}{H_I} e^{-2N} ~ \Omega_r^{1/2} ~ (\chi_{\calC}^2 H_0/c) \simeq 4 \times 10^8 l_{\rm Planck} 
\label{eq:achi}
\eeq
where for the second equality we have used the canonical value of $N \simeq 60$
and $H_i \simeq H_I$, which also yields $a_i \simeq 1.56 \times 10^{-53}$
and $H_I \sim 10^{10}$ GeV.
 The condition  $a_i \chi_{\calC}> l_{\rm Planck}$ requires $N<70$, close to the value found in \cite{DodelsonHui2003}. 

\begin{figure}[t]
\centering\includegraphics[width=0.8\linewidth]{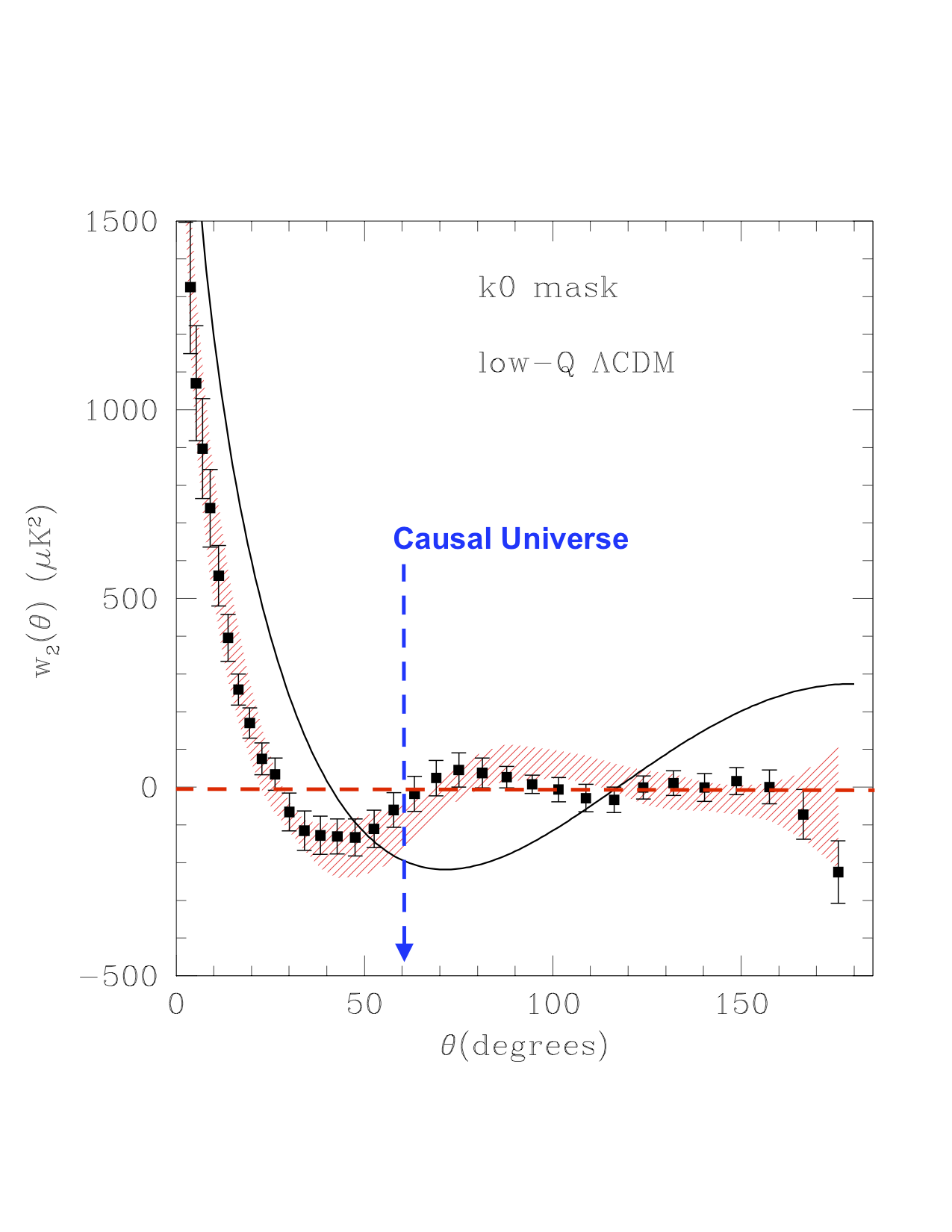}
\caption{Points with error-bars show t
Two-point correlation function of measured CMB temperature fluctuations in WMAP (points with errorbars) as a function of angular separation (adopted from Fig.12 in \cite{Gaztanaga2003}). There is no correlations above $\theta >60$ deg, which corresponds to the causal boundary (vertical dashed line) for $\Omega_\Lambda \simeq 0.69$. The black continuous line shows the  $\Lambda$CDM prediction for an infinite Universe. 
The shaded (red) region shows $\Lambda$CDM for a finite universe.
}
\label{Fig:CMB}
\end{figure}

\subsection{Implications for CMB}
\label{sec:CMB}

The (look-back) comoving distance to the surface of last scattering $a_* \simeq 9.2 \times 10^{-4}$ \cite{Planck2018} is
$\chi_{CMB} = \eta(1)- \eta(a_*) \simeq 3.145 ~ \frac{c}{H_0}$. This is shown as the horizontal dashed line in the bottom of Fig.\ref{Fig:conformal}. This is just slightly smaller than our estimate of the scale when the causal boundary re-enters the Horizon.
Thus, we would expect to see no correlations in the CMB on angular scales  $\theta > \theta_\calC$, where:

\beq
\theta_\calC \equiv 2 \arcsin{\frac{\chi_\calC /2}{\chi_{CMB}}} \simeq 60.1 ~\rm{deg.}
\label{eq:CMB}
\eeq
The lack of structure seen in the CMB on these large scales is one of the well known anomalies in the CMB data (e.g. see \cite{Copi2009,Schwarz2016} and references therein). 
Fig.\ref{Fig:CMB} (from \cite{Gaztanaga2003}) shows a comparison of the measured CMB temperature correlations (points with error-bars) with the $\Lambda$CDM prediction  for an infinite Universe (continuous line). There is a very clear discrepancy, which \cite{Copi2009} estimates to happen in only 0.025 per cent of the realizations of the infinite $\Lambda$CDM model. If we suppress the large scale modes above $\simeq 60$ deg in the $\Lambda$CDM simulations, the agreement is much better (shaded red area in Fig.\ref{Fig:CMB}). 


We can also predict $\Omega_\Lambda$ from the lack of CMB correlations. From Fig.\ref{Fig:CMB} we estimate $\theta_\calC = 62.2 \pm 4.0$ deg. to find (using Eq.\ref{eq:CMB} and Eq.\ref{Eq:rhoHchi}):
\beq
\theta_\calC = 62.2 \pm 4.0 ~\rm{deg.} \Rightarrow \Omega_\Lambda = 0.63 \pm 0.11
\eeq
 Note that the above estimate does not take into account the foreground ISW and lensing effects \cite{Fosalba,ISW}, which will typically  reduce $\theta_\calC$ slightly.
The value most used in the literature, $\theta_\calC \simeq 60$ deg., corresponds to $\Omega_\Lambda \simeq 0.70$.

Note that there are temperature differences on scales larger $\theta_\calC$, but they are not correlated, as expected in causality disconnected regions.
Nearby regions are connected which creates a smooth background across disconnected regions.

\section{Discussion and Conclusions}

$\Lambda$CDM in Eq.\ref{eq:Hubble} assumes that $\rho$ is constant everywhere at a fixed comoving time.
This requires acausal initial conditions \cite{Brandenberger} unless there is inflation, where a tiny homogeneous and causally connected patch, the Causal Universe $\chi_\calC$, was inflated to be very large today. Regions larger than  $\chi_\calC$ are out of causal contact. Here we require that 
test particles become free as we approach $\chi_\calC$.
This leads to Eq.\ref{eq:rhoH}, which is the main result in this paper.

If we ignore  the vacuum for now, this condition requires:
$\Lambda  = 8\pi G \rho_{\calC}$, where $\rho_\calC$ is the  matter and radiation inside 
$\chi_\calC$ (Eq.\ref{eq:rhoH4}). For an infinite Universe
($a_{\calC} \rightarrow \infty$) we have $\rho_{\calC} \Rightarrow 0$ which requires $\Lambda \Rightarrow 0$. This is also what we find in classical gravity with a $\Lambda$ term, because Hooke's term diverges at infinity (see Eq.\ref{eq:hooke}).  So  the  fact that $\rho_\Lambda \neq 0$  could indicates that  $\chi_\calC$ is not  infinite. Adding vacuum  $\rho_{\rm vac}$ does not change this as 
we have that $\Lambda  = 8\pi G \rho_{\calC} -  \rho_{\calC}$ so that $\rho_\Lambda = \Lambda/ 8\pi G + \rho_{\rm vac} = \rho_{\calC}$ turns out to be independent of $\rho_{\rm vac}$. Thus, whether the Causal Universe is finite or not, $\rho_{\rm vac}$ can not gravitate.

For constant vacuum ($\omega=-1$), we find  $\chi_\calC \simeq 3.15 c/H_0$ 
for  $\Omega_\calC= \Omega_\Lambda \simeq 0.69$. 
We can also estimate $\chi_\calC$ as $c/(a_i H_i)$ 
when inflation begins, see Eq.\ref{eq:aH}. After inflation  $\chi_\calC$ freezes out until it re-enters causality at $a_\calC \simeq 0.93$, close to now ($a=1$). This starts a new inflation (as $\rho_\Lambda = \rho_{\calC}> \rho_m$) which keeps the causal boundary frozen. 
Thus a finite  $\chi_\calC$  explains why $\rho_\Lambda \simeq 2.3\rho_m$ and
looking at Fig.\ref{Fig:horizon}, we argued that $a_\calC$ is the best time for observers like us to exist.

For $\omega =-1$, the measured value of $\Omega_\Lambda \simeq 0.70$ predicts that CMB temperature should not be correlated above $\theta> \theta_\calC \simeq 60$ deg, a prediction that matches observations
(see Fig.\ref{Fig:CMB}). One can also reverse the argument and use the lack of CMB correlations above
$\theta_\calC \simeq 60$ deg, to predict the size of $\chi_\calC  \simeq 2 \chi_{CMB} \sin{(\theta_\calC/2)}$. Together with condition $\rho_\calC=\rho_\Lambda$, this  provides a  prediction of $\Omega_\Lambda \simeq 0.70$, which is independent of other measurements for $\Omega_\Lambda$.

 If DE exists, we have shown that only the evolving component of DE is observable.
A universe with $\omega<-1$ violates our causality condition.
 In the limit of an infinite Universe with $\omega>-1$, we find that $\rho_\Lambda =\rho_{DE}$ (see Eq.\ref{eq:rhoDE}).
 But DE gives no clue as to why  $\rho_{DE} \simeq 2.3 \rho_{m}$ and can not explain  the lack of CMB correlations for $\theta>60$ deg. We apply Occam's razor to argue that there is no need for DE: measurements of cosmic acceleration and CMB can be explained by the finite size of our Causal Universe, as predicted by inflation.

\begin{acknowledgments}
I want to thank A.Alarcon, J.Barrow, C.Baugh, R.Brandenberger, G.Bernstein, M.Bruni, S.Dodelson,  E.Elizalde,  J.Frieman, M.Gatti, L.Hui, D.Huterer, A. Liddle, P.J.E. Peebles, R.Scoccimarro and S.Weinberg for their feedback.
This work has been supported by MINECO  grants CSD2007-00060 and AYA2015-71825, LACEGAL Marie Sklodowska-Curie grant No 734374 with ERDF funds from the European Union  Horizon 2020 Programme. IEEC is partially funded by the CERCA program of the Generalitat de Catalunya. 
\end{acknowledgments}







\end{document}